# Superconductivity in CaSn$_3$ single crystal with a AuCu$_3$-type structure

X. Luo[1†], D. F. Shao[1†], Q. L. Pei[1], J. Y. Song[1], L. Hu[1], Y. Y. Han[2], X. B. Zhu[1], W. H. Song[1], W. J. Lu[1*] and Y. P. Sun[1,2,3*]

[1] Key Laboratory of Materials Physics, Institute of Solid State Physics, Chinese Academy of Sciences, Hefei, 230031, China

[2] High Magnetic Field Laboratory, Chinese Academy of Sciences, Hefei, 230031, China

[3] Collaborative Innovation Center of Advanced Microstructures, Nanjing University, Nanjing, 210093, China

**Abstract**

We report the superconductivity of the CaSn$_3$ single crystal with a AuCu$_3$-type structure, namely cubic space group $P_{m\bar{3}m}$. The superconducting transition temperature $T_C$=4.2 K is determined by the magnetic susceptibility, electrical resistivity, and heat capacity measurements. The magnetization versus magnetic field (*M-H*) curve at low temperatures shows the typical-II superconducting behavior. The estimated lower and upper critical fields are about 125 Oe and 1.79 T, respectively. The penetration depth *λ(0)* and coherence length *ξ(0)* are calculated to be approximately 1147 nm and 136 nm by the Ginzburg-Landau equations. The estimated Sommerfeld coefficient of the normal state $\gamma_N$ is about 2.9 mJ/mol K$^2$. *ΔC/γ$_N$T$_C$* =1.13 and *λ$_{ep}$*=0.65 suggest that CaSn$_3$ single crystal is a weakly coupled superconductor. Electronic band structure calculations show a complex multi-sheet Fermi surface formed by three bands and a low density of states (DOS) at the Fermi level, which is consistent with the experimental results. Based on the analysis of electron phonon coupling of AX$_3$ compounds (A=Ca, La, and Y; X=Sn and Pb), we theoretically proposed a way to increase $T_C$ in the system.

[†]X. Luo and D. F. Shao contributed equally to this work.

*Author to whom correspondence should be addressed. Electronic mail:wjlu@issp.ac.cn and ypsun@issp.ac.cn

# I. Introduction

LnM$_3$ (Ln=Y and rare-earth elements, M=Pb, Tl, In, Ga, and Sn) compounds crystallized in a cubic AuCu$_3$-type structure have attracted much attention due to their interesting physical properties, such as superconductivity, heavy fermion behavior, quantum critical point and so on.[1,2] LnSn$_3$ compounds are particularly intriguing as they are found to have relatively higher superconducting transition temperature $T_C$. For instance, the $T_C$ of LaSn$_3$ is 6.5 K and that of YSn$_3$ is 7.0 K, whereas Pb, Tl, In, and Ga related compounds have lower $T_C$.[3-6] On the other hand, some LnSn$_3$ compounds show additional interesting physical phenomena. For example, PrSn$_3$ has been reported to be a heavy fermion compound with an antiferromagnetic order at $T_N$=8.6 K and CeSn$_3$ is categorized as a dense Kondo compound with valence fluctuation.[7,8] However, the isostructural ASn$_3$ (A=alkaline earth metal) compounds are seldomly explored; the ion-ratio mismatch between the alkaline earth metal ions and Sn one, it is difficult to crystallize into a AuCu$_3$-type structure. For instance, BaSn$_3$ and SrSn$_3$ with superconducting transition temperature $T_C$=2.4 K and 5.4 K distort into hexagonal and rhombohedra structures, respectively.[9,10] Different from the distorted BaSn$_3$ and SrSn$_3$ compounds, CaSn$_3$ with smaller Ca ion can be crystalized into a AuCu$_3$-type structure and it has the same structure with CaPb$_3$ and CaTl$_3$ superconductors.[11-13] Although the polycrystalline CaSn$_3$ was synthetized thirty years ago, the physical properties of CaSn$_3$ have been rarely reported because the CaSn$_3$ is a decomposition product of highly air-sensitive. To the best of our knowledge, the de Has-van Aplhen effect has only been observed in polycrystalline CaSn$_3$ powder.[11] In order to get a deeper insight into the physical properties of CaSn$_3$ single crystal, further research is needed.

Herein, we have successfully grown the bulk single crystals of CaSn$_3$ for the first time and discovered that it shows superconductivity with $T_C$=4.2 K. We have performed electrical resistivity, magnetic susceptibility, and heat capacity measurements to determine the superconducting parameters of CaSn$_3$ single crystal. Furthermore, electronic band structure calculations suggest a complex, multi-sheet Fermi surface formed by three bands. The calculated low density of states (DOS) at the Fermi level is almost consistent with the value derived from the small electron specific coefficient measured. Finally, based on the analysis of electron phonon coupling (EPC) of AX$_3$ compounds (A=Ca, La, and Y; X=Sn and Pb), we theoretically proposed a way to increase $T_C$ in the system.

## II. Experimental details

Single crystalline specimens of CaSn$_3$ were prepared through Sn-self flux. Ca pieces (99.99 %, Alfa Aesar) and Sn powders (99.99 %, Alfa Aesar) with mole ratio 1:28 were weighted and loaded into a 2 mL alumina crucible, which was sealed in an evacuated quartz tube. All were done in an Ar-filled glove box. The sealed quartz tubes were slowly heated to 800 $^o$C for 24 hours, and dwelled for 24 hours, then slowly cooled down to 270 $^o$C with 2 $^o$C/h. Finally, the quartz tubes were inverted and quickly spun into a centrifuge to remove the excess Sn flux. More detail can be found in the Supporting information. Rectangular single crystals with shining surface were observed. The size was about 1.5*1.5*1 mm$^3$. The single crystals were air-sensitive and thus they were kept inside the glove box until characterization. Such handling was necessary to avoid decomposition. Powder X-ray diffraction (XRD) patterns were taken with Cu $K_{α1}$ radiation (λ=0.15406 nm) using a PANalytical X'pert diffractometer at room temperature. Heat capacity and electrical transport properties were measured using the Quantum Design physical properties measurement system (PPMS-9T) and magnetic properties were performed by the magnetic property measurement system (MPMS-XL5). The electrical transport measurements were performed by a four-probe method to eliminate the contact resistance. The measurement of specific heat was carried out by a heat-pulse relaxation method on PPMS-9T. Electronic structures were obtained from first-principles density functional theory in the generalized gradient approximation (GGA) according to the Perdew-Burke-Ernzerhof.[14] The QUANTUM-ESPRESSO package was used with ultrasoft pseudopotentials generated by Garrity, Bennett, Rabe, and Vanderbilt (GBRV).[15,16] The energy cut off for the plane-wave basis set is 40 Ry. Brillouin zone sampling is performed on the Monkhorst-Pack (MP) mesh of 16×16×16.[17]

## III. Results and Discussion

CaSn$_3$ shows a cubic structure with space group $P_{m\bar{3}m}$, as present in Fig. 1 (a). Ca ions set in the corner and six Sn ions are octahedrally coordinated without atoms inside. Figure 1(b) shows the powder X-ray of CaSn$_3$ single crystals which were ground in a glove box. The sample powder was sealed by Kapton film during the measurement to protect it from oxidation. As shown in Fig. 1 (b), it presents a cubic structure with little Sn-flux impurity, which may be due to the decomposition of $CaSn_3$ powder during the grinding and data collection. The high intensity in the low degree zone is the background from the Kapton film. The lattice constant *a* calculated from indexes is about 4.742 Å, which matches well with the previously reported value.[11]

Figure 1 (c) shows the resistivity dependence of temperature of $CaSn_3$ single crystal from 300 K to 2 K. The electrical resistivity data show a metallic behavior ($\frac{d\rho}{dT} > 0$) with the large residual resistivity ratio (*RRR*): $\frac{\rho_{300 K}}{\rho_{5 K}} \sim 73$, which indicates the $CaSn_3$ single crystal is of high quality. A clear superconducting transition can be observed at 4.2 K. We applied the Fermi-liquid model $\rho(T) = \rho_0 + AT^2$, where $\rho_0$ and *A* are the residual resistivity and a constant, respectively, to the curve below 20 K as shown in the inset of Fig. 1 (c). The model analysis yielded the residual resistivity $\rho_0$ of 1.3 $\mu\Omega\ cm$ and the coefficient *A* of 0.00227 $\mu\Omega\ cm/K^2$, indicating Fermi-liquid-like behavior for $CaSn_3$ single crystal.

The *ρ* versus *T* curve of $CaSn_3$ single crystal was further analyzed over the whole temperature range by the Bloch-Grüneisen-Mott (BGM) model, which is expressed as[18]

$$\rho_{BGM}(T) = \rho_0 + 4\mathcal{R}T\left(\frac{T}{\Theta_R}\right)^4 \int_0^{\Theta_R/T} \frac{x^5}{(e^x-1)(1-e^{-x})} dx \quad , \quad (1)$$

where $\rho_0$ is the residual resistivity and $\mathcal{R}$ is a constant. The second term represents contributions from the electron-phonon interaction and $\Theta_R$ is the Debye temperature.[19] Least-squares fit of the *ρ(T)* data by Eqs. (1) for 5 K≤*T*≤300 K is shown in Fig.1 (c). Fig. 1S shows the fitting data at low temperature. The parameters were estimated to $\rho_0$=1.5 $\mu\Omega\ cm$, $\mathcal{R}$=0.17 $\mu\Omega\ cm/K$, and $\Theta_R$=211 K.

Figure 1 (d) presents the temperature dependence of magnetization with the zero-field and field cooling (ZFC and FC) modes under applied magnetic field 10 Oe. The bulk superconducting transition temperature $T_C$=4.2 K can be clearly seen in the left inset of Fig. 1 (d), which is consistent with the resistivity result. The magnetization as a function of magnetic field at 2.5 K is shown in the right inset of Fig. 1 (d). The clear hysteresis is observed, which indicates a typical type-II superconducting behavior of $CaSn_3$ single crystal. The little element Sn may be left on the $CaSn_3$ single crystal surface, we compared the magnetization as a function of magnetic field curves of $CaSn_3$ single crystal and element Sn at *T*=3.6 K, as shown in Fig. 2S. The $CaSn_3$ single crystal shows a typical type-II superconductor, which is different from the element Sn with type-I superconducting behavior.

In order to gain further insights into the superconducting state, the magneto-resistivity and the magnetic field dependence of magnetization at various temperatures were performed. Figure 2 (a) shows the resistivity under different magnetic fields. $T_C$ at different magnetic field is determined at a 50 % decrease from the normal-state resistivity value, and transition width is taken as the temperature interval between 10% and 90 % of the transition. Figure 2 (b) presents the $H_{C2}$-$T$ phase diagram of the superconducting state of CaSn$_3$ single crystal. We estimate the upper critical field, $H_{C2}(0)$, from the Werthamer-Helfand-Hohenberg (WHH) expression[20]

$$\mu_0 H_{C2}(0) = -0.693 (dH_{C2}/dT)_{T=T_C} T_C \ . \quad (2)$$

A nearly linear relationship is obtained in the measured temperature range, which leads to a $\mu_0 H_{c2}(0)$ value of 1.79 T. The value of $\mu_0 H_{c2}(0)$ is lower than the Pauli-limiting field

$$\mu_0 H^{Pauli} = 1.24 k_B T_C / \mu_B \ . \quad (3)$$

Expected within the same weak-coupling BCS theory, it is about 7.6 T for $T_C$=4.2 K.[21] The empirical formula:

$$H_{C2}(T) = H_{C2}(0) \left[ 1 - \left( \frac{T}{T_C} \right)^{\frac{3}{2}} \right]^{\frac{3}{2}} \ , \quad (4)$$

which is usually used to calculate the upper critical field for a variety of intermetallic and oxide superconductors.[22,23] The value of upper field is about $\mu_0 H_{c2}(0)$=1.86 T, which agrees well with the $\mu_0 H_{c2}(0)$ obtained from WHH method. The upper critical field value $\mu_0 H_{c2}(0)$ can be used to estimate the Ginzburg-Landau (GL) coherence length $\xi(0) = \sqrt{\frac{\Phi_0}{2\pi H_{C2}(0)}} = 136$ Å, where $\Phi_0 = hc/2e$ is the flux quantum.

Figure 2 (c) shows the magnetization as a function of magnetic field at different temperatures below $T_C$. $\mu_0 H_{c1}(0)$ as a function of $T$ is present in the Fig. 2 (d). The lower critical field is analyzed using the equation:

$$H_{C1}(T) = H_{C1}(0) \left[ 1 - \left( \frac{T}{T_C} \right)^2 \right] \ . \quad (5)$$

The value of $\mu_0 H_{C1}(0)$ is about 125 Oe. In addition, we may estimate the Ginzburg-Landau (GL) parameter $\kappa = 1/\sqrt{2} [H_{C2}(0)/H_C(0)]$ ~8.4, where the thermodynamic critical field $H_C(0)=(H_{C1}(0)H_{C2}(0))^{1/2}$~0.15 T, and the magnetic penetration depth $\lambda(0)$ is found to be 1147 Å from $\lambda(0)=\kappa\xi(0)$.[24]

The temperature dependence of specific heat $C_P$ also presents more information about the superconducting properties. Figure 3 (a) shows the temperature dependence of the heat capacity $C_P$ of CaSn$_3$ single crystal at $H$=0 and 5 T. At $H$=0 T, the specific heat data show the bulk superconducting feature as indicated by the clear jump of $C_P$ around $T_C$. At $H$=5 T, the specific heat jump of superconductivity in CaSn$_3$ single crystal is completely suppressed. $C_P/T$ varies almost linearly with $T^2$ above $T_C$. The Sommerfeld constant $\gamma_N$ is obtained from the fit $C_P/T=\gamma_N+\beta T^2$ ($\gamma_N$ and $\beta$ are $T$-independent coefficients), where $\gamma_N$ is the normal-state electronic contribution and $\beta$ is the lattice contribution to the specific heat. The fitting results yield $\gamma_N$ =2.9 mJ/mol K$^2$ and $\beta$=1.45 mJ/mol K$^4$. The Debye temperature $\Theta_D$ can be determined from the coefficient of the $T^3$ term $\beta = N(12/5)\pi^4 R\Theta_D^{-3}$, where $R$=8.314 J/mol K and $N$=4 for CaSn$_3$ single crystal. $\Theta_D$ is about 175 K, which is little lower than that of YSn$_3$.[6] The Debye temperature $\Theta_D$ obtained from the low temperature heat capacity $C_P$ was slightly lower than the Debye temperature $\Theta_R$=211 K derived from the resistivity data because the $\Theta_R$ obtained from the resistivity was averaged over the whole temperature range. It is known that the ratio $\Delta C_P/\gamma_N T_C$ can be used to estimate the coupling strength.[25] As shown in the inset of Fig. 3, $\Delta C_P/\gamma_N T_C$ of CaSn$_3$ single crystal from the $C_P$ data is about 1.13, which is slightly lower than the expected BCS value with $\Delta C_P/\gamma_N T_C$ ~1.43 for superconductors within the weak-coupling limit. This suggests that the CaSn$_3$ single crystal is a weakly coupled superconductor. An estimation of the strength of the EPC can be derived from the McMillan formula: [26,27]

$$\lambda_{ep} = \frac{\mu^* \ln\left(\frac{1.45 T_C}{\Theta_D}\right)-1.04}{1.04+\ln(\frac{1.45 T_C}{\Theta_D})(1-0.62\mu^*)} \qquad . \tag{6}$$

The EPC constant $\lambda_{ep}$ is estimated to be 0.65, assuming the Coulomb pseudopotential $\mu^*$=0.1, which supports the weakly coupling scenario. The determined superconducting parameters of CaSn$_3$ single crystal are listed in Table **I**.

Figure 4 shows the calculated electronic structure of CaSn$_3$ based on the density functional theory calculations. Figure 4(a) presents the band structure in the vicinity of Fermi energy ($E_F$). According to the calculated band structure, CaSn$_3$ is a three-dimensional (3D) metal. The bands near $E_F$ are predominately contributed by $p$ electrons of Sn ions (Fig. 4 (b)). Three bands with large dispersion cross the Fermi level, forming the complex 3D Fermi surfaces (shown in Figs. 4 (c)-(e)). From the calculated DOS (Fig. 4 (b)), one can notice that $E_F$ locates at a valley of DOS, leading to a small $N(E_F)$ of 0.82 states/eV. The resulted $\gamma_{cal} = \frac{1}{3}\pi^2 k_B^2 N(E_F)$ = 1.93 mJ/mol K$^2$. Such value is lower than that of our measured CaSn$_3$ ($\gamma_{exp}$ = 2.9 mJ/mol K$^2$) obtained from the specific heat data. The enhancement of $\gamma$ should be attributed to EPC with the factor of $\frac{\gamma_{exp}}{\gamma_{cal}} = 1 + \lambda_{ep}$, where $\lambda_{ep}$ is derived to be ~0.5, which is qualitatively in agreement with the weakly coupling scenario proposed above.

In order to understand the superconductivity of CaSn$_3$ single crystal, herein, we compare CaSn$_3$ superconductor with other AuCu$_3$ type alloys. As we know, researchers usually want to find a general rule to guide the exploration of the superconductors with higher $T_C$ in the AuCu$_3$ type alloys. For example, the oscillation of $T_C$ with respect to numbers of valence electrons $n$ was suggested.[3, 13] On the other hand, a recent study shows that $T_C$ monotonically increases with when the lattice constant $a$ decrease for AX$_3$ (A=Y, La; X=Sn, Pb) compounds in AuCu$_3$ structure.[6] However, we found those rules are be applied to other AuCu$_3$-type alloys. For example, our present CaSn$_3$ single crystal has smaller lattice parameter $a$ but much lower $T_C$ than that of LaSn$_3$ ($T_C$~6.5 K).[4] Moreover, the oscillation of $T_C$ with respect to numbers of valence electrons $n$ seems to be only effective for the AX$_3$ alloys in which the X atoms are from the same period.[3] Therefore, we should reconsider the superconductivity in AX$_3$ (X is metal elements in IV$_A$ group) with AuCu$_3$ structure. Figure 5 shows the calculated DOS for the reported ASn$_3$ and APb$_3$ (A=Ca, Y, and La) compounds, which are of good agreement with previous reports.[27] The shapes of DOS curves near $E_F$ for ASn$_3$ (A=Ca, Y, and La) are very similar except for the $E_F$ that decided by the number of valence electrons $n$. YSn$_3$ and LaSn$_3$, which can be seen as one electron doped CaSn$_3$, have higher $E_F$ locates at a small DOS peak (Fig. 5 (a)). Due to the large atomic number of Pb ion, the spin-orbit coupling is considered in the calculation of APb$_3$. One can notice that the spin-orbit coupling makes the peaks of DOS sharper (Fig. 5 (b)).

We summarize the $N(E_F)$, lattice parameter $a$, and $T_C$ of the reported ASn$_3$ and APb$_3$ (A=Ca, Y, La) superconductors in Fig. 6. The simple monotonic relationships between $T_C$ and the lattice parameter *a* or numbers of valence electrons *n* cannot be obtained. As we known, the EPC strength $\lambda_{ep}$ for a material can be qualitatively expressed as:

$$\lambda_{ep} = \sum_\alpha \frac{\langle I_\alpha^2 \rangle N_\alpha(E_F)}{M_\alpha \langle \omega_\alpha^2 \rangle}, \qquad (7)$$

where the $\langle I_\alpha^2 \rangle$, $M_\alpha$, and $\langle \omega_\alpha^2 \rangle$ are the mean square EPC matrix element averaged over Fermi surface, atomic mass, and averaged squared phonon frequency of the $\alpha$th atom in the unit cell, respectively.[29] Since the $N(E_F)$ of $AX_3$ (X=Sn and Pb) is predominately contributed by X, we can qualitatively propose that the total EPC is mainly decided by

$$\lambda_X = \frac{\langle I_X^2 \rangle N(E_F)}{M_X \langle \omega_X^2 \rangle}, \qquad (8)$$

where X=Sn and Pb. Therefore, though La has a much larger atomic mass, when X is fixed, the $T_C$ differences between $YX_3$ and $LaX_3$ are small (less than 1 K), since the $N(E_F)$ of those two are close. Meanwhile, when X is fixed, $YX_3$ and $LaX_3$ have higher $N(E_F)$ than that of $CaX_3$, which leads to a higher $T_C$ of $YX_3$ and $LaX_3$. On the other hand, when X is changed from Sn to Pb, the EPC was strongly suppressed by the large increase of mass in the denominator of Eq. (8). For example, $YPb_3$ and $LaPb_3$ have the $N(E_F)$ three times larger than that of $CaSn_3$. However, the $T_C$ of the three compounds are on the same level (~4 K). The $N(E_F)$ of $CaPb_3$ is 50% higher than that of $CaSn_3$. However, the $T_C$ of $CaPb_3$ ($T_C \sim 1$ K) is only one fourth of that of $CaSn_3$. According to the above analysis we can choose the suitable doping elements to increase $N(E_F)$ or decrease the mass of X atom to enhance the EPC and hence increase $T_C$. Moreover, compared to other cubic superconductors, such as $MgCNi_3$ ($T_C \sim 7$ K) with $N(E_F)$ of ~5 states/eV [30] and $La_3Al$ ($T_C \sim 6$ K) with $N(E_F)$ of ~4 states/eV[31], $CaSn_3$ with superconductivity at the liquid helium temperature has a very small $N(E_F)$, which is unexpected. This indicates a large $\langle I^2 \rangle$ in the system. If $AX_3$ compounds with a lighter X atom in $IV_A$ group (i.e., $AGe_3$, $ASi_3$) can be obtained, the superconductivity with a much higher $T_C$ might probably be achieved.

## IV. Conclusion

In summary, we have investigated the physical properties of $CaSn_3$ single crystals. The electrical resistivity, magnetic susceptibility, and specific heat confirm that the superconductivity transition temperature of $CaSn_3$ is 4.2 K. Both $\Delta C_e/\gamma_N T_C$ and $\lambda_{ep}$ value magnitudes indicate that the $CaSn_3$ is a weakly coupled superconductor. Furthermore, electronic band structure calculations are consistent with experimental data and show the complex and multi-sheet Fermi surfaces. Based on the analysis of electron phonon coupling of AX$_3$ compounds (A=Ca, La, and Y; X=Sn, Pb), we theoretically concluded a way to increase $T_C$ in the system.

# Acknowledgments

This work was supported by the Joint Funds of the National Natural Science Foundation of China and the Chinese Academy of Sciences' Large-Scale Scientific Facility under contracts (U1432139, U1232139), the National Nature Science Foundation of China under contracts (11304320, 51171177), the National Key Basic Research under contract 2011CBA00111, and the National Nature Science Foundation of Anhui Province under contract 1508085ME103. The authors thank Dr. Chen Sun for her assistance in editing the manuscript.

**Table I:** Superconducting parameters of CaSn$_3$ single crystal.

| Parameters | Units | CaSn$_3$ |
|---|---|---|
| $T_C$ | $K$ | 4.2 |
| $\rho_0$ | $\mu\Omega\,cm$ | 1.3 |
| $A$ | $\mu\Omega\,cm/K^2$ | $2.27 \times 10^{-3}$ |
| $RRR$ | | 73 |
| $\mu_0 H_{c1}(0)$ | $Oe$ | 125 |
| $\mu_0 H_{c2}(0)$ | $T$ | 1.79 |
| $\mu_0 H_C(0)$ | $T$ | 0.15 |
| $\mu_0 H^{pauli}$ | $T$ | 7.6 |
| $\xi(0)$ | $\text{Å}$ | 136 |
| $\lambda(0)$ | $\text{Å}$ | 1147 |
| $\kappa(0)$ | | 8.44 |
| $\gamma(0)$ | $\frac{mJ}{mol\,K^2}$ | 2.9 |
| $\beta$ | $\frac{mJ}{mol\,K^4}$ | 1.45 |
| $\frac{\Delta C}{\gamma T_C}$ | | 1.13 |
| $\Theta_D$ | $K$ | 175 |
| $\Theta_R$ | $K$ | 211 |
| $\lambda_{ep}$ | | 0.65 |

# Figure captions:

**Fig. 1 (Color online):** (a) Cubic perovskite crystal structure of $CaSn_3$; (b) Powder XRD patterns at room temperature for the crushed $CaSn_3$ single crystals. The vertical red bars and blue ones stand for the position of Brag peaks of $CaSn_3$ and Sn in PDF card, respectively. The inset presents the picture of $CaSn_3$ single crystal used for this study; (c) The temperature dependence of the resistivity of $CaSn_3$ single crystal. The red solid line is the fitting by using the Bloch-Grüneisun-Mott (BGM) model. Inset shows an enlarged view of the plot in a low temperature range ($T<30$ K). The solid line is the fitting to the data using from Fermi-liquid model; (d) ZFC and FC magnetic susceptibility of $CaSn_3$ single crystal measured at $H=10$ Oe. Inset of (i) shows the temperature dependence of magnetization in a large scale; (ii) presents the magnetic field dependence of magnetization at $T=2.5$ K.

**Fig. 2 (Color online):** (a) Temperature dependence of $\rho(T)/\rho(5\ K)$ under magnetic field of 0, 500, 1000, 2000, 3000, 4000, and 5000 Oe from right to left, respectively; (b) The upper critical field $\mu_0H_{C2}$ as a function of $T_C$. Solid lines are fits to Eq. (2) and (4), respectively; (c) The magnetic field dependence of magnetization at different temperatures; (d) The lower critical field $\mu_0H_{C1}$ dependence of $T_C$. The solid line is fit to Eq. (3).

**Fig. 3 (Color online):** The main panel shows the heat capacity in 0 T and 5 T magnetic fields. The blue line shows the heat capacity data fitting with the equation $C_P/T=\gamma_N+\beta T^2$. The inset shows that the heat capacity jumps around $T_C$.

**Fig. 4 (Color online):** (a) Band structure of $CaSn_3$. The bands crossing $E_F$ are colored; (b) DOS of $CaSn_3$; (c), (d), and (e) are the Fermi surface from the bands colored with red, blue, and purple in (a), respectively.

**Fig. 5 (Color online):** DOS of (a) $ASn_3$ and (b) $APb_3$ for A=Ca, Y, and La.

**Fig. 6 (Color online):** Lattice parameter $a$ dependence of $T_C$ for the reported $AX_3$ (X is metal elements in $IV_A$ group) superconductors (data from Ref. 13).

**Figure 1:**

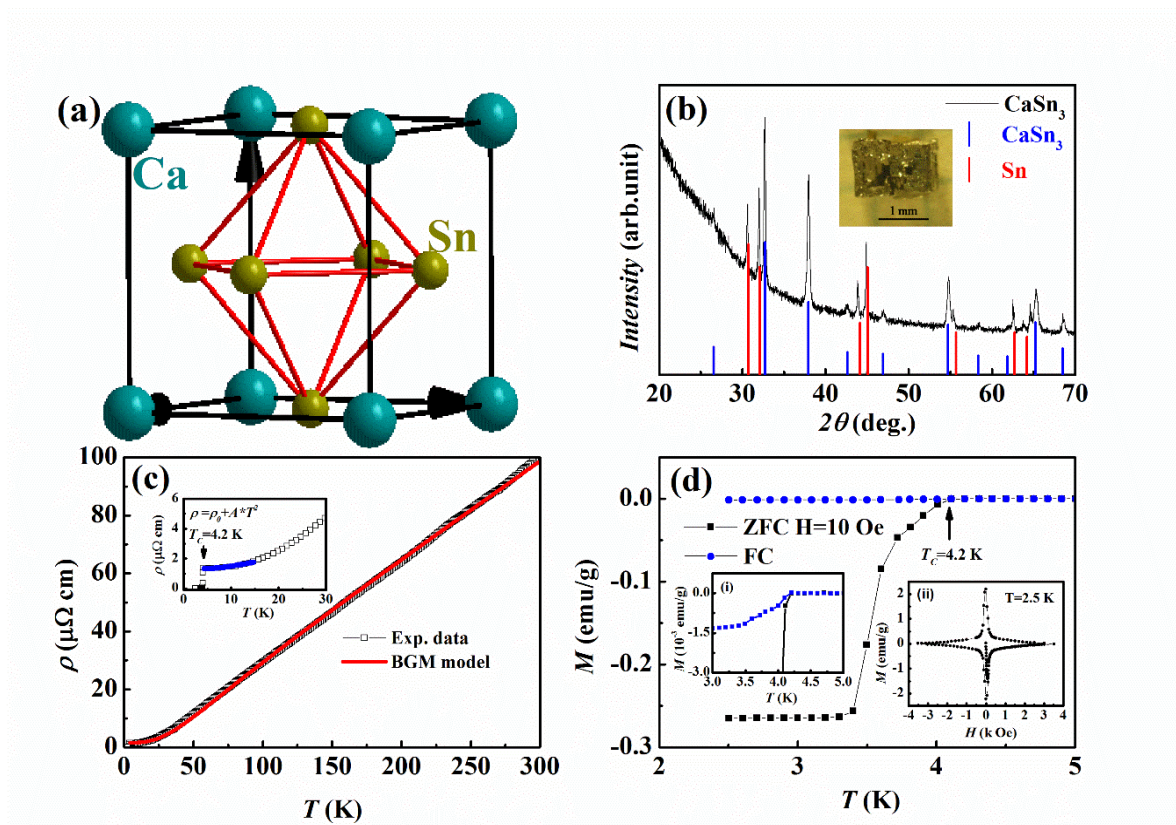

**Fig. 1 X. Luo *et al.***

**Figure 2:**

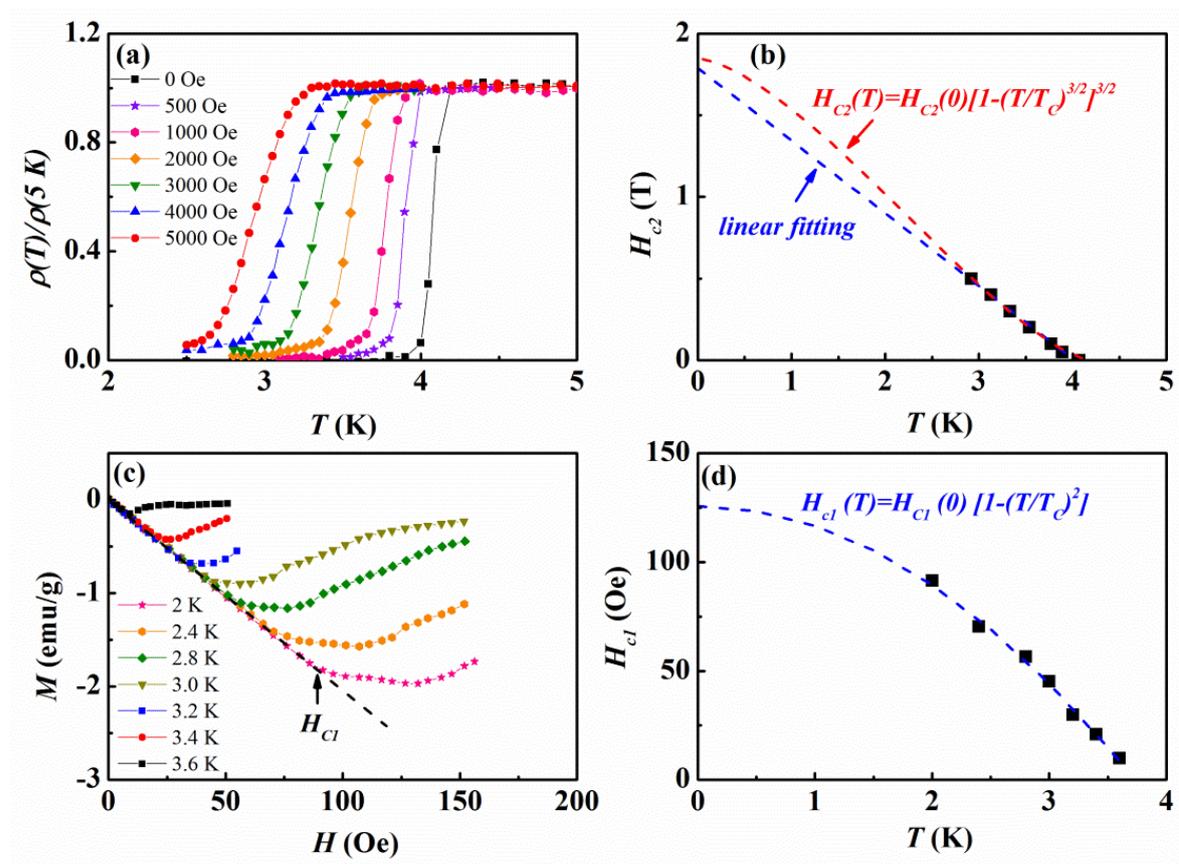

**Fig.2 X. Luo *et al.***

**Figure 3:**

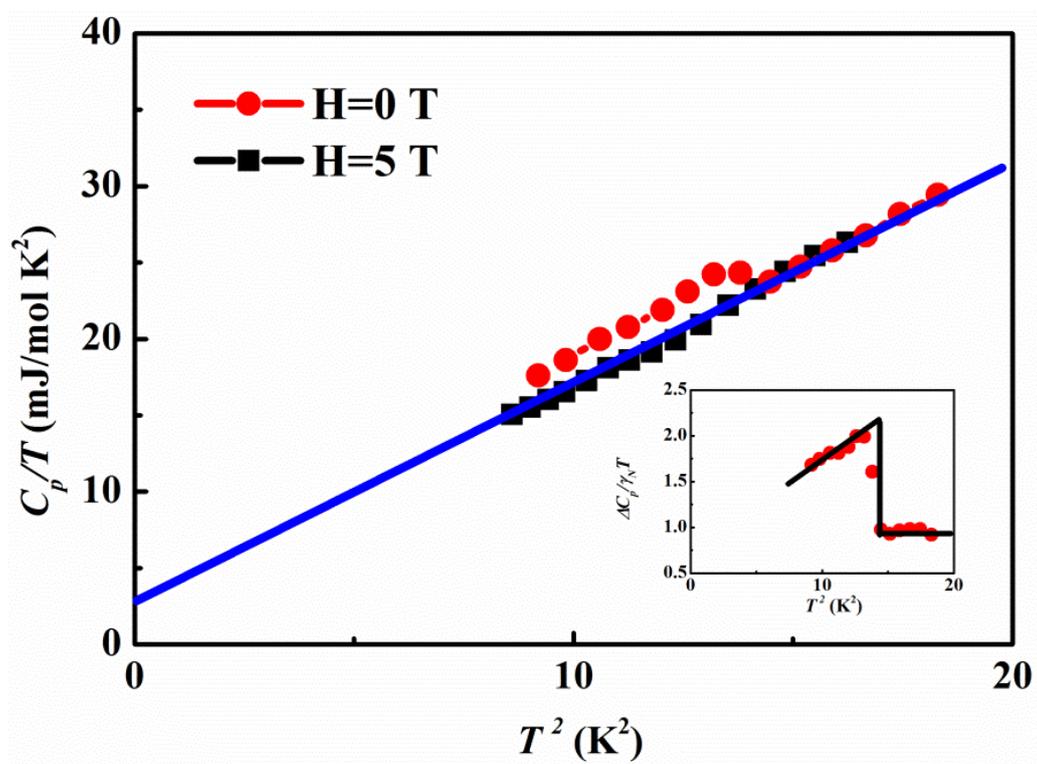

**Fig. 3 X. Luo *et al.***

**Figure 4:**

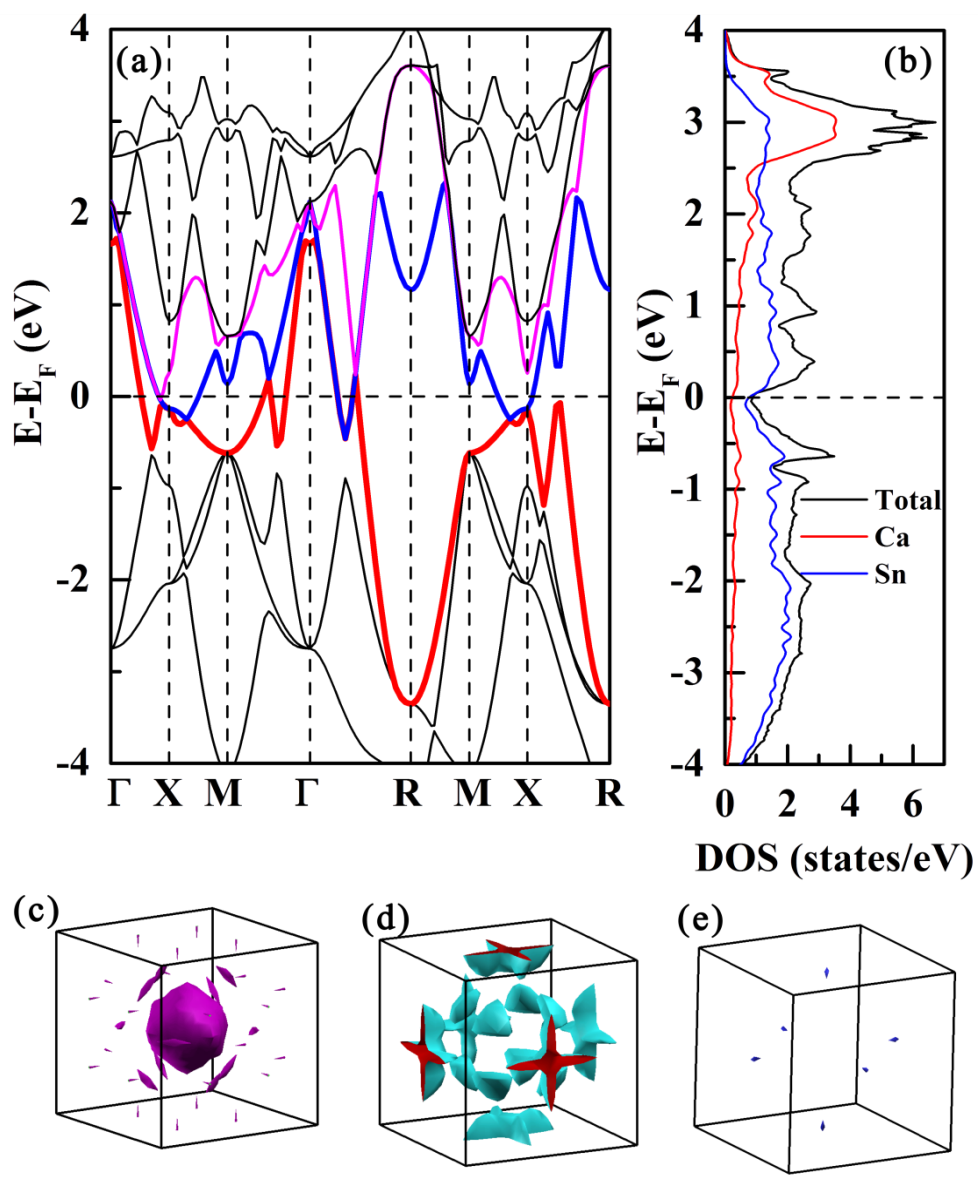

**Fig.4** X. Luo *et al.*

**Figure 5:**

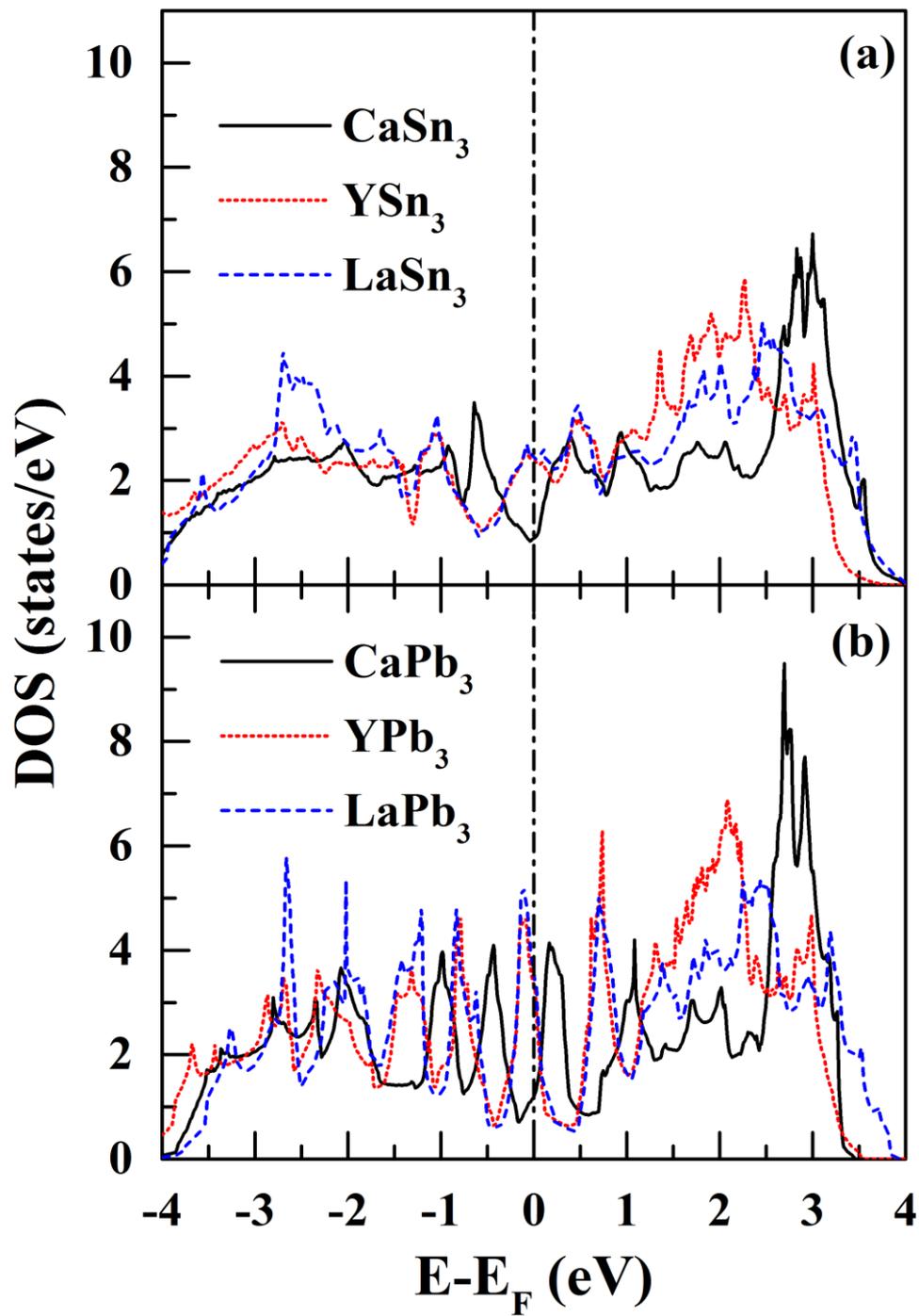

**Fig.5 X. Luo *et al.***

**Figure 6:**

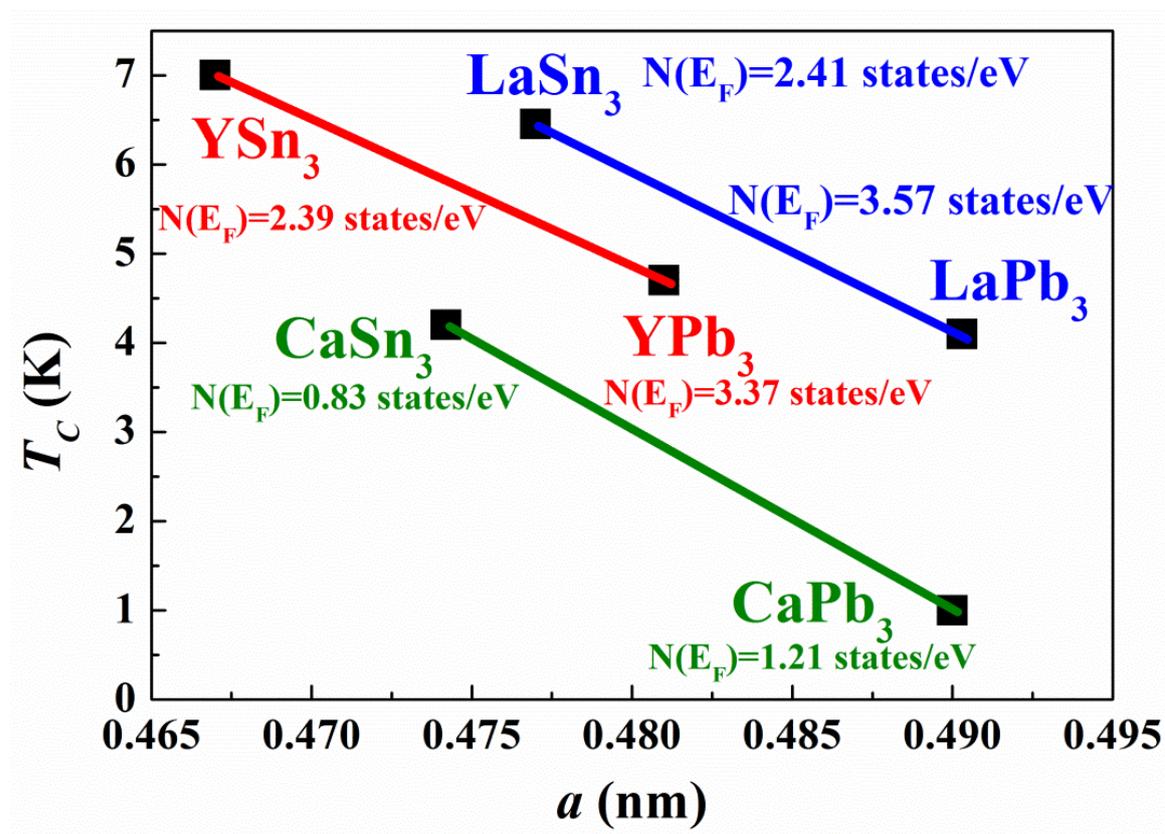

# Supporting Information

## Superconductivity in CaSn$_3$ single crystal with a AuCu$_3$-type structure

X. Luo[1†], D. F. Shao[1†], Q. L. Pei[1], J. Y. Song[1], L. Hu[1], Y. Y. Han[2], X. B. Zhu[1], W. H. Song[1], W. J. Lu[1*] and Y. P. Sun[1,2,3*]

[1] Key Laboratory of Materials Physics, Institute of Solid State Physics, Chinese Academy of Sciences, Hefei, 230031, China

[2] High Magnetic Field Laboratory, Chinese Academy of Sciences, Hefei, 230031, China

[3] Collaborative Innovation Center of Advanced Microstructures, Nanjing University, Nanjing, 210093, China

## Abstract

We report the superconductivity of the CaSn$_3$ single crystal with a AuCu$_3$-type structure, namely cubic space group $P_{m\bar{3}m}$. The superconducting transition temperature $T_C$=4.2 K is determined by the magnetic susceptibility, electrical resistivity, and heat capacity measurements. The magnetization versus magnetic field (*M-H*) curve at low temperatures shows the typical-II superconducting behavior. The estimated lower and upper critical fields are about 125 Oe and 1.79 T, respectively. The penetration depth *λ(0)* and coherence length *ξ(0)* are calculated to be approximately 1147 nm and 136 nm by the Ginzburg-Landau equations. The estimated Sommerfeld coefficient of the normal state $\gamma_N$ is about 2.9 mJ/mol K$^2$. *ΔC/γ$_N$T$_C$* =1.13 and *λ$_{ep}$*=0.65 suggest that CaSn$_3$ single crystal is a weakly coupled superconductor. Electronic band structure calculations show a complex multi-sheet Fermi surface formed by three bands and a low density of states (DOS) at the Fermi level, which is consistent with the experimental results. Based on the analysis of electron phonon coupling of AX$_3$ compounds (A=Ca, La, and Y; X=Sn and Pb), we theoretically proposed a way to increase $T_C$ in the system.

Electronic mail: wjlu@issp.ac.cn and ypsun@issp.ac.cn

**More experimental details:**

The used crystals were cubic or rectangle shape, we decanted the crystals is about 270 ºC, which is higher than the melting point of element Sn (231 ºC) and decanting speed is very fast and can reach 1800 round/second within 10 seconds. We did the polishing before doing the measurements, so just little element Sn may be left on the surface.

**Figures:**

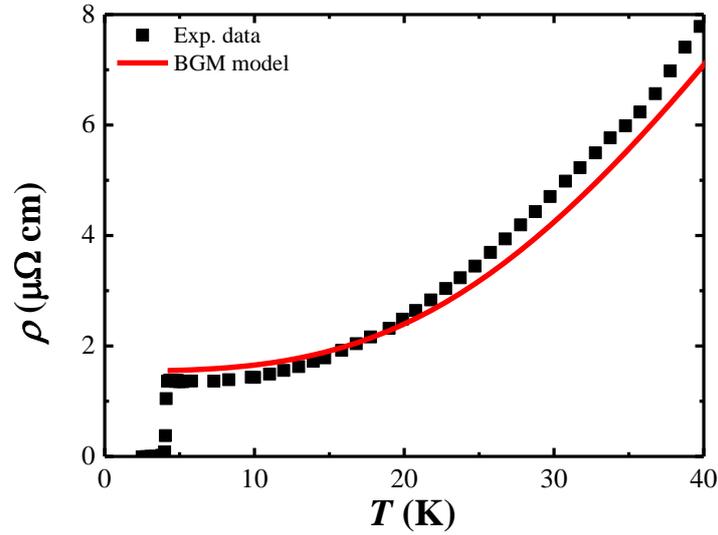

Fig. S1: The fitting result using the BGM model is shown at the low temperature.

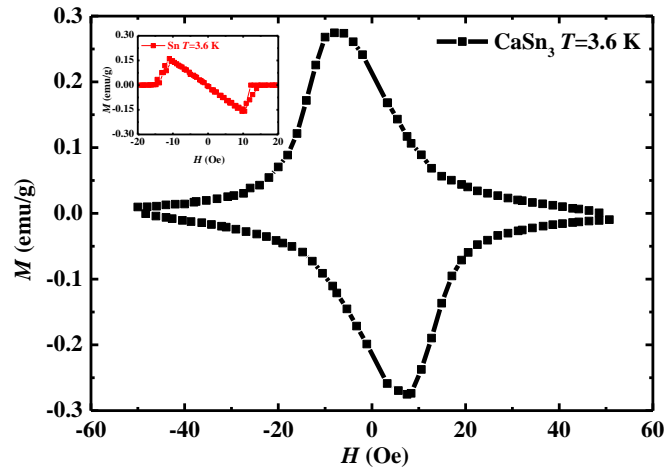

Fig. S2: The comparation of $M(H)$ between the CaSn$_3$ single crystal and element Sn at $T$=3.6 K.